\documentclass[aps,prl,twocolumn,showpacs,floatfix]{revtex4}
\usepackage{amsmath}
\usepackage{graphicx,epsfig,psfrag}
\usepackage{amssymb}
\begin{document}

\title{Strongly enhanced thermal transport in a lightly doped Mott insulator at low temperature}
\author{V. Zlati\'c$^{1,2}$}
\author{ J. K. Freericks$^{2}$}
\affiliation{
$^{1}$Institute of Physics, Zagreb POB 304, Croatia\\
$^{2}$ Department of Physics, Georgetown University, Washington D.C., 20057, USA
}
\begin{abstract}
We show how a lightly doped Mott insulator has hugely enhanced electronic thermal transport at low temperature.  
It displays universal behavior independent of the interaction strength when the carriers can be treated 
as nondegenerate fermions and a nonuniversal ``crossover'' region where the
Lorenz number grows to large values, while still maintaining a large thermoelectric figure-of-merit.
The electron dynamics are described by the Falicov-Kimball model which is 
solved for arbitrary large on-site correlation with a dynamical mean-field theory algorithm  on a Bethe lattice.   
We show how these results are generic for lightly doped Mott insulators as long as the renormalized Fermi 
liquid scale is pushed to very low temperature and the system is not magnetically ordered.
\end{abstract}

\maketitle

\paragraph{Introduction}
\label{zlatic:sec:1}
Thermoelectric materials are attracting significant attention, because of their potential for various 
power generation or refrigeration applications which involve so-called green technologies. 
The mass application of the thermoelectric devices is hampered by their low efficiency 
and the aim of current research is to produce materials  
with better thermoelectric  conversion efficiency.
At high temperatures, the efforts are directed towards nanostructured 
semiconductors~\cite{boetner_2006} with reduced thermal conductivity. 
At  low temperatures, the focus is on the materials with strongly correlated electrons, 
like  Kondo insulators and systems with a Mott-Hubbard gap to enhance the thermopower in metallic systems. 

Here, we present a theory for the charge and thermal transport in a slightly doped 
Mott insulator that is  described by the Falicov-Kimball model.  
The exact solution is obtained from dynamical mean-field theory (DMFT) and it
shows that for large correlation and small doping the figure-of-merit  
is unusually  large. This behavior is expected to also hold in more general doped Mott insulators, 
like those described by the Hubbard model if they are sufficiently frustrated away from magnetic order and
if the operating temperature lies well above the renormalized Fermi temperature. 
Even the presence of magnetic order is unlikely to strongly modify this effect, since the magnetic order 
rarely has a large effect on charge and heat transport in strongly correlated materials\cite{wilhelm_jaccard}.
The numerics, however, requires quite high precision to determine the transport for such low
doping values, which is why this problem can only be solved within the Falicov-Kimball model
with current state-of-the-art methods.

\paragraph{Theoretical description} 

To discuss this thermoelectric phenomena, we use transport equations which express 
the charge and the internal energy current densities,  ${\bf J}({\bf x})$ and ${\bf J}_{\cal E}({\bf x})$, 
in terms of the generalized forces~\cite{luttinger.64}. 
The coefficients of these generalized forces, $N_{ij}(T)$, are given by 
various current-current correlation functions which have to be calculated for the model at hand.  
The electrical conductivity $\sigma$, the Seebeck coefficient (or thermopower) $\alpha$,  and the thermal 
conductivity $\kappa_e$ are then obtained as  
\begin{eqnarray}
\sigma(T)
&=&
e^2 N_{11}(T)
~,
                      \label{eq: sigma}
\\
\alpha(T)
&=&
\left(\frac{k_B}{e}\right)
\frac{N_{12}(T)}{T N_{11}(T)}~, 
                           \label{eq: thermopower}                             
\\
\kappa_e(T)
&=&
\left(\frac{k_B}{e}\right)^2\frac{\sigma(T)}{T} D(T)
~,
                           \label{eq: thermal_conductance}                             
\end{eqnarray} 
where $D=N_{22}/N_{11} - N^2_{12}/N^2_{11}$ gives the effective Lorenz number 
${\cal L}(T)=\kappa_e/\sigma T=(e/k_B)^2 D(T)/T^2$. 
The   dimensionless figure-of-merit of a particular thermoelectric material is 
$ZT=\alpha^2\sigma T/\kappa$, where  $\kappa=\kappa_e+\kappa_{ph}$ is the 
overall thermal conductivity due to the electronic and  the lattice degrees of freedom.  
The electronic figure-of-merit can be expressed as $ZT=\alpha^2/{\cal L}$ and 
an efficient thermoelectric material has $ZT > 1$. 
We do not consider the phonon contribution to the thermal conductivity further here.

We next show how to exactly evaluate the transport coefficients for 
the spinless Falicov-Kimball model~\cite{falicov_kimball_1969} 
with a large on-site Coulomb interaction.  
The Hamiltonian on an infinite-coordination ($Z\rightarrow\infty$ with $Z$ the coordination number 
in this equation only and not to be confused with the figure of merit $ZT$) 
Bethe lattice is
\begin{equation}
\mathcal{H}=-\frac{t^*}{\sqrt{Z}}\sum_{\langle ij\rangle}(c^{\dag}_ic^{}_j+h.c.)-(\mu+\frac{U}{2})\sum_ic^{\dag}_ic^{}_i+U\sum_ic^\dag_ic^{}_iw_i
~, 
				\label{eq: hamtotal}
\end{equation}
where $c^{\dag}_i$ ($c^{}_i$) creates (destroys) a conduction electron at site $i$, $t^*$ 
is the renormalized hopping, $\mu$ is the chemical potential shifted so that $\mu=0$ corresponds to half-filling, 
$U$ is the interaction between localized and itinerant electrons, and $w_i$ is a classical variable 
equal to 0 if a localized electron is not at site $i$ and equal to 1 if the localized electron is at site $i$. 
The average filling of the localized electrons is fixed at 1/2, so that the system is in a Mott-insulating state 
when $U$ is larger than $2t^*$ and $\mu=0$.  
The sum over $i$ and $j$ in the kinetic-energy term is over nearest neighbor pairs and $h.c.$ denotes the hermitian conjugate.
This model is appropriate for doped Mott insulators where the dopants provide carriers without producing 
impurity levels in the gap, as is often seen in many semiconductors.  

\begin{figure} 
\includegraphics[width=\linewidth,clip]{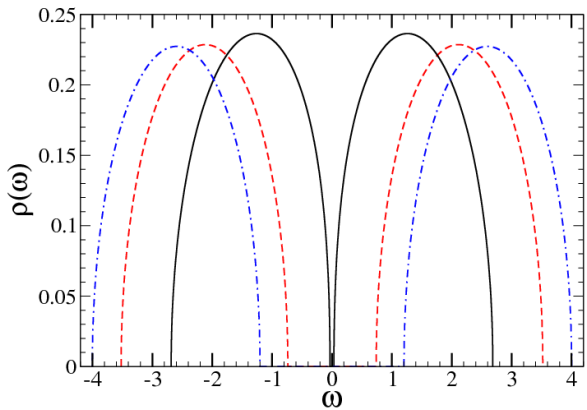}
\caption{ (Color online) 
The density of states plotted as a function of energy  for 
$U=2.2$ (full line),  $U=4$ (dashed line), and  $U=5$ (dashed-dotted line). 
The lower and the upper gap-edges are at  $\omega_{-}(U)$ and $\omega_{+}(U)$, 
respectively. 
\label{figure1} 
}
\end{figure}

To calculate the thermal transport, we need to evaluate a series of Kubo formulas for 
the relevant transport coefficients.  This is greatly simplified because the Falicov-Kimball model satisfies the 
Jonson-Mahan theorem~\cite{jonson_mahan_1990,freerics_zlatic_MJ}.
Thus, the transport coefficients can be written as~\cite{freericks_zlatic_fk_review}, 
\begin{eqnarray}
N_{mn}(T)       
=
\int_{-\infty}^{\infty}d\omega~
\left(-\frac{\partial f(\omega)}{\partial \omega}\right)\ \omega^{m+n-2} \ 
\Lambda^{}_{tr}(\omega)~, 
                           \label{eq: transport_integrals}                             
\end{eqnarray}
with the transport function given by 
\begin{eqnarray}
\Lambda^{}_{tr}(\omega)
=
\frac{4}{3\pi^2}
\int d\epsilon ~
\rho^{}_{0}(\epsilon) (4t^{*2}-\epsilon^2)
[\rm Im ~ G^{}_{}(\epsilon,\omega)]^2~,
                      \label{eq: lambda_tr}
\end{eqnarray}
where the noninteracting density of states is 
\begin{eqnarray}
\rho_{0}(\epsilon)
=
\frac{1}{2\pi t^*}\sqrt{4t^{*2}-\epsilon^2}~, 
\label{eq: rho_2D}
\end{eqnarray} 
and $G^{}_{}(\epsilon,\omega)$ is the ``band-energy-dependent'' Green's function 
($\epsilon$ is the noninteracting band energy)
which we calculate 
within DMFT~\cite{freericks.2004}. 
(One should note that there is no momentum on a Bethe lattice, and instead, 
one should think of the band-energy as analogous to the independent variable of momentum.) 
The integral in Eq.~\eqref{eq: lambda_tr}  
can be performed exactly, resulting in
\begin{equation}
\Lambda_{tr}(\omega)=\frac{1}{3\pi^2}{\rm Im}^2[G(\omega)]\frac{|G(\omega)|^2-3}{|G(\omega)|^2-1}.
\end{equation}
The local Green's function, $G(\omega)=\int d\epsilon \rho_0(\epsilon) G(\epsilon,\omega)$ 
satisfies a simple cubic equation~\cite{vandongen}, 
which allows for the numerics to be carried out to high precision ({\it i.e.}, 
there is no self-consistent iterative algorithm needed to solve the problem). 
In this equation, and in the following, we set $t^*=1$ as our energy unit.
Note that because the density of states at the band edge grows like the square root of frequency, 
the transport density of states is linear near the band edge as one goes into the band and vanishes as one goes into the gap.
Similar results are obtained for a three-dimensional cubic lattice (within DMFT), 
as has been checked for a few points in parameter space.

\paragraph{Numerical results}
The energy dependence of the  renormalized density of states 
$[\rho(\omega)=-{\rm Im}G(\omega)/\pi]$ of a Mott insulator is shown 
in Fig.~\ref{figure1} for several values of $U$, with $\mu=0$. 
The density of states for the Falicov-Kimball model does not depend 
on temperature, and since the cubic equation for $G(\omega)$ depends 
only on $\omega+\mu$, $U$, and the average filling of the localized electrons, 
the density of states for a lightly doped Mott insulator is identical to that of 
the Mott insulator, except the origin is shifted, as the chemical potential changes. 
For $U>2$, the DOS is split into a lower and upper Hubbard band of width $W$ 
and with a separation between the maxima approximately equal to $U$. 
The shape of the Hubbard bands is nearly independent of $U$ (for $U > 2$) and 
only the gap, which extends from  $\omega_{-}(U)$ to $\omega_{+}(U)$, increases with $U$.  

\begin{figure}
\includegraphics[width=\linewidth,clip]{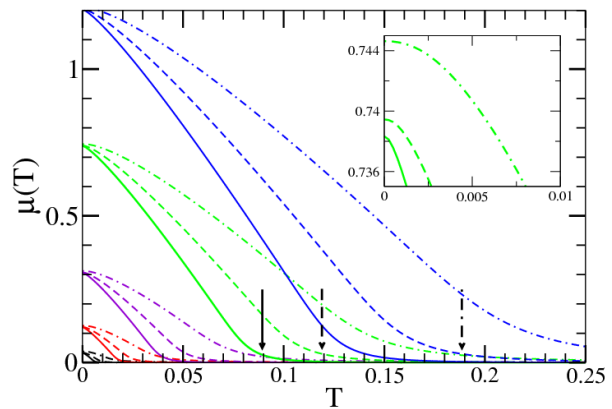}
\caption{{ (Color online)} 
Chemical potential plotted as a function of temperature.  
The black, red, purple, green, and blue curves are calculated for $U=$2.2, 2.5, 3,  4, 
and 5, respectively. The full, dashed, and dashed-dotted curves correspond to the concentrations 
of $n_c=10^{-6}$, $10^{-5}$, and $10^{-4}$  electrons above half filling.    
The characteristic temperature $T_\mu$, obtained for $U=4$, 
is indicated by the full, dashed and dashed-dotted arrows. 
The inset show the low-temperature behavior obtained for $U=$4.
\label{figure2}}
\end{figure}
The temperature dependence of the chemical potential, obtained for various values of $U$ and 
several (small) values of  the upper Hubbard band filling $n_c$, is shown in Fig.~\ref{figure2}. (The total conduction electron filling is just $1/2+n_c$.)
At zero doping, the system is electron-hole symmetric and the chemical potential is 
in the middle of the gap, $\mu(T)=0$.  
At finite electron doping, the zero-temperature chemical potential is in the upper Hubbard band, 
just above the band edge, as can be seen by comparing $\mu(0)$ 
in Fig.~\ref{figure2} with $\omega_{+}(U)$ in  Fig.~\ref{figure1}.
For $n_c\ll 1$, the values of $\mu(0)$ are approximately given by 
$\omega_{+}(U)\approx (U-W)/2$.  
The  low-temperature behavior of $\mu(T)$ is demonstrated in the inset 
of Fig.~\ref{figure2}. The data show that after an initial parabolic decrease,  
$\mu(T)$ is nearly linear up to a characteristic temperature $T_\mu$. 
For $T > T_\mu$,  the decrease of $\mu(T)$ slows down, because excitations across 
the gap restore the electron-hole symmetry and the chemical potential  
approaches the high-temperature limit, $\mu(T)\to 0$.
For a given $U$, the characteristic temperature $T_\mu$ increases with $n_c$. 
This is shown in  Fig.~\ref{figure2}  by the full, dashed, and dashed-dotted arrows 
which indicate $T_\mu$ obtained for $U=$ 4 and $n_c=10^{-6}$, $10^{-5}$, and $10^{-4}$, respectively.  

The observed behavior follows from the fact that $\mu(T)$ is close to the 
bottom of the upper Hubbard band at low temperatures and that the excitations across the gap
establish a symmetric state at high temperatures, where $\mu\approx 0$. 
Thus, for a given $n_c$, an increase of $U$ shifts $\mu(0)$ and $T_\mu$ to higher values 
and translates $\mu(T)$ upwards, as shown in Fig.~\ref{figure2}. 
If we increase $n_c$ keeping $U$ constant,  a higher temperature is needed to restore the 
electron-hole symmetry, so that $T_\mu$ increases with $n_c$.

\begin{figure}
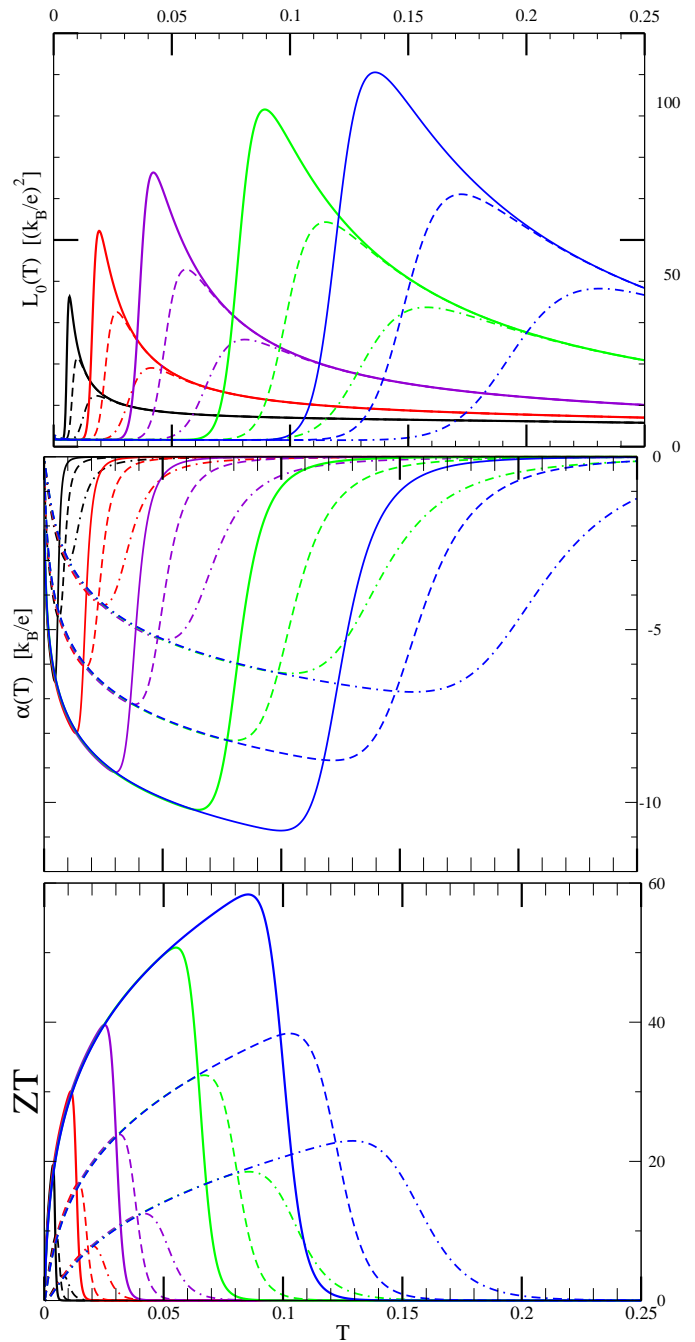

\includegraphics[width=8.9cm,clip]{Lorenz+U}
\vskip -0.1 cm
\includegraphics[width=\linewidth,clip]{TEP+U}
\vskip -0.25 cm
\includegraphics[width=8.7cm,clip]{ZT+U}
\caption{ {(Color online)}  
Lorenz number in units of $[k_B/e]^2$, thermopower in units of $[k_B/e]$, and thermoelectric figure of merit
$ZT$ are plotted as a function of temperature for the same parameters as in Fig.~\ref{figure2}.
\label{figure3}
}
\end{figure}

To calculate the transport integrals in Eq.~(\ref{eq: transport_integrals})
we adjust $\mu(T)$ to yield the target filling of particles
and measure the energy  at each temperature with respect to $\mu(T)$. 
Since neither $\rho(\omega)$ nor $\Lambda_{tr}(\omega)$ 
change their shape,   the transport 
coefficients  in Eq.~(\ref{eq: transport_integrals})  are easy to compute.  
At high temperatures, $T \geq T_\mu(U)$,  the behavior  is universal, 
{\it i.e.},  the transport coefficients are independent of $n_c$,  for a given $U$.
This is revealed most clearly by the effective Lorenz number, ${\cal L} (T)$,  
plotted in the uppermost panel of Fig.~\ref{figure3}. 
The curves ${\cal L} (T)$ obtained for a given  $U$ and various $n_c$ merge 
for $T\geq T_\mu$, due to the fact that at such high temperatures $\mu\simeq 0$ 
and all the systems we are concerned with become (nearly) electron-hole symmetric. 

At the lowest temperatures, $T\leq T_1$,  the Lorenz number is given by the usual value, 
${\cal L} (T)\simeq {\cal L}_0=(\pi^2/3)(k_B/e)^2 $, expected from the Wiedemann-Franz 
law for degenerate fermions. 
The characteristic temperature $T_1$ depends on the distance of the $T=0$ 
chemical potential from the upper Hubbard band edge $\mu(0)-\omega_+(U)$.  
For very low doping, 
this temperature is physically irrelevant and we do not show it on the figure. 
Another type of universality sets in when the chemical potential is in the gap and 
the lower Hubbard band starts affecting  the transport. 
Figure \ref{figure3} shows that, for a given $U$, the Lorenz number becomes 
independent of $n_c$ for a wide range of temperature $T_1\ll T <T_0$, where 
$T_0$ depends on the size of the Mott-Hubbard gap.  
Here, the Fermi gas is nondegenerate and  the Wiedemann-Franz law holds with 
the smaller classical value  ${\cal L}_0=2(k_B/e)^2$. 

Curiously, for $T_1\ll T<T_0$, 
the thermopower and the figure of merit also assume 
universal forms, independent of $U$, which can be seen by the overlapping 
curves in Fig.~\ref{figure3}.  
This low-temperature universality arises because 
the shape of $\Lambda_{tr}(\omega)$  around $\omega=\mu$  
does not depend strongly on $U$.
The universality is lost  for $T\geq T_0$, when the excitations across the gap become important.  

For $T_0\leq T \leq T_\mu $, there is a crossover from the low- to the high-temperature 
regime. Here, the transport coefficients vary in a non-universal way, which is also revealed 
most clearly by the effective Lorenz number. 
It grows rapidly for $T\geq  T_0(n_c) $, attains a maximum at about $T\simeq T_\mu$, 
and then decays slowly. 
Well above  $ T_\mu$,  where $\kappa$ and $\sigma$ assume their high-temperature 
universal forms, ${\cal L} (T)$ is independent of $n_c$. 

We now consider in more detail the results for thermopower and the figure-of-merit,  shown in Fig.~\ref{figure3}. 
At very low temperatures, the slope of $\alpha(T)$ is determined by the logarithmic derivative 
$d \ln{ \Lambda_{tr}}/d\omega|_{\omega=\mu}$ which is large for low doping,  
since $\mu(T)$ is close to the band edge. 
Here,  $\alpha(T)$ grows rapidly and $\alpha/T$ increases as $n_c$ decreases 
but, for a given $n_c$, it does not depend on $U$.   
At higher temperatures, the presence of the Mott-Hubbard gap leads to novel features. 
Unlike in ordinary semiconductors, $\alpha(T)$ continues to grow when the temperature renormalization 
brings $\mu(T)$ below the bottom of the upper Hubbard band.  
When that happens, the states below the chemical potential do not contribute to $N_{12}$ 
and  $\alpha(T)$ grows to very large values for large $U$ and $T\leq T_0$. 
Eventually, for $T_0 < T \simeq T_\mu$, the system becomes nearly electron-hole symmetric 
and $\alpha(T)$ drops to very small values.  

The figure-of-merit becomes very large in the proximity of the metal-insulator transition. 
For $U\geq 20$ and doping which gives less than $10^{-19}$ 
conduction electrons ($n_c\approx 10^{-4}$), we find  $ZT > 200$!
The  maximum of $ZT$ occurs at a  temperature at which  the  Wiedeman-Franz law holds for nondegenerate fermions
and the thermal conductivity is very low, so that the thermal current  due to 
any additional degrees of freedom will greatly reduce the thermoelectric efficiency.
However, the temperature range in which $ZT>1$  increases  with $U$ and $n_c$. 
For larger doping, a relatively large $ZT$ is found in a broad temperature range, even for moderate $U$.  
By tuning the concentration of conduction electrons, one can find a temperature window in which  
${\cal L}_0 \ll {\cal L} (T) $ and  $ZT\gg 1 $. Here, the electronic thermal conductivity is 
not small and the large $ZT$ values should be physically relevant, if one were to include phonon effects to the thermal conductivity.
This occurs, typically for temperatures of the order of $T_\mu$.

\paragraph{Summary and conclusions}
We presented a theory for the charge and heat transport of a "bad metal"  described by the Falicov-Kimball model. 
Using DMFT, we show that the transport coefficients of a lightly doped Mott insulator  exhibit universal features 
at low temperatures, $T_1\ll T\leq T_0$, and at high temperatures, $T\geq T_\mu$. 
In the crossover region, $T_0 \leq T\leq T_\mu$, the effective Lorenz number 
becomes very large and the Wiedemann-Franz law is not obeyed. 
The thermopower and the figure of merit grow to very large values close to $T\approx T_0$. 
In this region, the system might be optimal for use in thermoelectric devices.

The low-temperature thermoelectric properties are enhanced because the 
renormalized transport density of states of a lightly doped Mott insulator is very asymmetric close to the gap. 
As the temperature increases, the chemical potential drops below the bottom of the conduction 
band, such that the asymmetry of the electron and hole currents becomes even more pronounced, 
and the thermopower is further enhanced. 
At about $T\simeq T_\mu$, the chemical potential approaches the center of the gap, 
the system nearly acquires electron-hole symmetry, and the thermopower drops to small values. 
We could not obtain similar effects with non-interacting electrons or in a Fermi liquid.

The $ZT$  of a strongly correlated system with just a few electrons in the upper Hubbard band 
turns out to be surprisingly large. This finding is not restricted to a Bethe lattice; 
large $ZT$  is obtained for a three-$d$ cubic lattice as well.  We point out that 
the maximum of $ZT$ occurs at temperatures at which the Wiedemann-Franz law holds for nondegenerate fermions
and the thermal conductivity is low, so that the thermoelectric efficiency of a real system 
described by our model would be greatly affected by other degrees of freedom that can transport heat.  
However, for large $U$ and moderate $n_c$, there is a broad temperature interval, 
$T_0 \leq T \leq T_\mu$,  in which $ZT$ is moderately large, even though 
the electronic thermal conductivity and ${\cal L}$  are not small and in this regime, phonons will not affect the $ZT$ values as much.

One might ask whether such features are generic for all Mott insulators, or specific to the Falicov-Kimball model. 
It turns out, the Hubbard model will share this behavior for small enough doping and large $U$.  This is because in this regime 
the renormalized Fermi temperature is strongly reduced towards zero, so a Fermi liquid won't form in this interesting 
regime where the thermoelectic effects can be so large, and the Hubbard model acts similar to the Falicov-Kimbal model.  
(It is still possible that a too large phonon thermal conductivity, that magnetic order which modifies the transport density 
of states, or localization effects due to disorder in lightly doped systems will reduce this effect, but there should be 
a regime where it will still be able to be seen~\cite{semiconductor}.) We hope that this work will inspire experimental groups to look for 
this kind of phenomena as a new route towards low-temperature thermoelectric refrigeration.
\acknowledgments
This work is supported by  the NSF grant No. DMR-1006605.  
V.Z. acknowledges support by Croatian MZOS Grant No.0035-0352843-2849
J.K.F. is also supported by the McDevitt bequest at Georgetown University.


\begin{thebibliography}{99.}
\bibitem{semiconductor} In a conventional semiconductor, localization effects can be more important than correlation effects if the localization length is less than the mean-free-path due to electron-electron 
correlations.  This is less likely to occur in Mott insulators that have stronger electron-electron correlations.
\end{thebibliography}
\end{document}